\documentclass[amsmath,amssymb,amsbsy,prl,twocolumn,superscriptaddress,longbibliography,floatfix]{revtex4-1}
\usepackage{amsfonts} 
\usepackage{amsmath} 
\usepackage{amssymb}
\usepackage{blindtext}
\usepackage{lineno}
\usepackage{cancel}
\usepackage{ulem}
\usepackage{bm}
\usepackage[usenames, dvipsnames]{color} 
\usepackage{dcolumn}
\usepackage{epic} 
\usepackage{epsfig}
\usepackage{eucal} 
\usepackage{esdiff}
\usepackage{graphicx}
\usepackage[breaklinks,colorlinks = true,linkcolor = magenta,urlcolor=magenta,citecolor=red]{hyperref}
\usepackage{mathrsfs}
\usepackage{soul}
\usepackage{subfigure}
\usepackage{textcomp}
\usepackage{times}
\usepackage{verbatim}
\usepackage{wrapfig} 
\usepackage{xy} 
\usepackage{lineno}
\usepackage{float}
\usepackage{import}
\newcommand{\pkg}{\mathrm{P}_{\!\!\! _{K_{\scriptscriptstyle
        \mathrm{G}}}}}
	\newcommand{\kg}{{K_{\scriptscriptstyle \mathrm{G}}}}

\begin{document}
\title{Many-body Chaos in Thermalised Fluids}
\author{Sugan D. Murugan} 
\email{sugan.murugan@icts.res.in}

\affiliation{International Centre for Theoretical Sciences, Tata Institute of Fundamental Research, Bengaluru 560089, India} 
\author{Dheeraj Kumar}
\email{dheeraj.kumar@espci.fr} 
\affiliation{PMMH, CNRS, ESPCI Paris,
Universit\'e PSL, Sorbonne Universit\'e, Universit\'e de Paris, 75005 Paris, France} 
\affiliation{International Centre for Theoretical Sciences, Tata Institute of Fundamental Research, Bengaluru 560089, India}
 \author{Subhro Bhattacharjee} 
 \email{subhro@icts.res.in} 
\affiliation{International Centre for Theoretical Sciences, Tata Institute of Fundamental Research, Bengaluru 560089, India} 
\author{Samriddhi Sankar Ray} 
\email{samriddhisankarray@gmail.com}
\affiliation{International Centre for Theoretical Sciences, Tata Institute of Fundamental Research, Bengaluru 560089, India}
\begin{abstract} 
Linking thermodynamic variables like temperature $T$ and the measure of chaos,
the Lyapunov exponents $\lambda$, is a question of fundamental importance in
many-body systems. By using nonlinear fluid equations in one and three
dimensions, we show that in thermalised flows $\lambda \propto \sqrt{T}$, in
agreement with results from frustrated spin systems. This suggests an
underlying universality and provides evidence for recent conjectures on the
thermal scaling of $\lambda$. We also reconcile seemingly disparate
effects---equilibration on one hand and pushing systems out-of-equilibrium on
the other---of many-body chaos by relating $\lambda$ to $T$ through the
dynamical structures of the flow.
\end{abstract}
\date{}
\maketitle

Many-body chaos is the key mechanism to explain the fundamental
basis---\textit{thermalisation} and \textit{equilibration}---of statistical
physics. However, there are equally important  examples in nature, such as turbulence, 
where chaos plays a role that is seemingly {\it
opposite} from the {\it settling down} through thermalisation and
equilibration of several many-body systems. This contrast becomes stark if we
argue in terms of the celebrated \textit{butterfly effect}
~\cite{lorenz1963deterministic,lorenz1996essence,lorenz2000butterfly,hilborn2004sea}:
While the amplification of the \textit{wingbeat} results in complex dynamical
macroscopic structures in driven-dissipative systems (e.g., a turbulent fluid),
the same amplification leads to a loss of memory of initial conditions,
resulting in ergodic behaviour and eventual thermalisation or equilibration,
in Hamiltonian many-body systems.  How then do we reconcile these two
apparently disparate roles of many-body chaos?  

An important piece of the answer lies in investigating the spatio-temporal
aspects (the Lyapunov exponent $\lambda$ and butterfly speed $v_{\mathrm{B}}$)
of many-body chaos in fluids to reveal its connection with macroscopic
(thermodynamic) characterisation of the system.  This provides for
comparisons of length and time scales of chaos and thermalisation, on the one
hand, and the non-linear dynamic structures of the fluid-velocity field on the
other.  

Characterisations of chaos and its connection with transport and hydrodynamics are recent  
in the context of both classical and quantum many-body systems like 
unfrustrated and frustrated~\cite{Das2018,PhysRevLett.121.250602,bilitewski2020classical,ruidas2020many} magnets, strongly
correlated field theories
\cite{Blake2016,Blake2017,Gu2017,Lucas2017,Werman2017,Patel2017,Patel1844,kitaev,PhysRevB.95.134302})
and  field theories of black-holes~\cite{Shenker2014,Cotler2017}. A common feature responsible
for the unconventional signatures of chaos in many of these systems seems to
originate from a large set of strongly coupled, dynamic, low energy modes arising  
from competing interactions. This is similar a 
turbulent fluid where the triadic interactions of velocity (Fourier) modes across several
decades lead to strong couplings resulting in, e.g., scale-by-scale energy transfers
\cite{kraichnan_71,orszag_70}.
    
These studies have been facilitated by the development of quantum
\textit{out-of-time commutators}
(OTOCs)~\cite{Larkin1969,kitaev,Maldacena2016,Das2018,Roberts2015a,Dora2017,Aleiner2016}
and their classical counterpart the  \textit{decorrelator}
\cite{Das2018,PhysRevLett.121.250602} which measure how two  \textit{very}
nearly identical copies of a system decorrelate spatio-temporally. 
The classical decorrelators are 
invaluable for understanding the butterfly
effect~\cite{lorenz1963deterministic,lorenz1996essence,lorenz2000butterfly,hilborn2004sea}
in non-integrable, chaotic, classical many-body systems through the measurement
of $\lambda$ and $v_{\mathrm{B}}$. Since by construction, these OTOCs or
decorrelators provide a unified framework to bridge thermodynamic variables
(e.g., temperature $T$) with the butterfly effect, they are a unique
prescription to connect many-body chaos with the foundations of statistical
approaches in  \textit{both} classical and
quantum many-body systems.  The most striking example of this is that while for
quantum systems, $\lambda \le T/\hbar$, limiting the rate of
scrambling~\cite{Maldacena2016}), the analogous \textit{conjecture} for
classical systems is $\lambda \propto \sqrt{T}$ at low
temperatures~\cite{Maldacena2016,Kurchan2016}.  

In this Letter,  by using a model of \textit{thermalised
fluids}, we derive $\lambda \propto \sqrt{T}$ 
and demonstrate a possible universality of  many-body chaos without an
apparent (weakly interacting) quasi-particle description, and hence a Kinetic
Theory.  Interestingly, we show how decorrelators \textit{sense}
the emergent dynamical structures of the fluid velocity field, providing an
elegant way to bridge the ideas of many-body chaos with foundational
principles of statistical physics: Thermalisation, equilibration and
ergodicity. 


For classical systems, recent understanding of
spatio-temporal chaos through decorrelators stems primarily from spin systems~\cite{PhysRevLett.121.250602,bilitewski2020classical,ruidas2020many} . However,  these ideas have
not been applied for the most ubiquitous of chaotic, nonlinear,  systems: 
Turbulent flows. This is because, unlike spin-systems,  turbulent flows,
governed by the viscous Navier-Stokes equation, are an example of a
driven-dissipative system \textit{without} a Hamiltonian or a statistical
physics description in terms of thermodynamic variables.  Therefore, we look for variations of the
Navier-Stokes equation which, whilst preserving the same non-linearity, nevertheless has a 
a Hamiltonian structure, resulting in a chaotic, \textit{thermalised} fluid.

\begin{figure}
	\centering
	\includegraphics[scale=0.8]{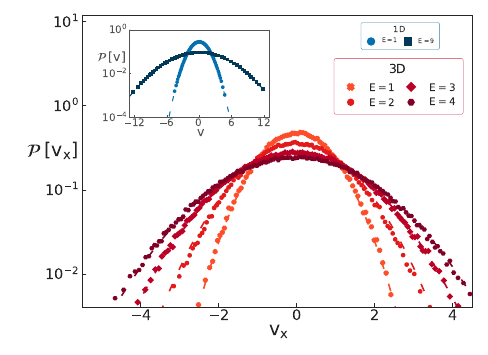}
	\caption{Probability distribution
	functions of the $x$-component of the thermalised velocity field from Galerkin-truncated 3D Euler and (inset) 1D Burgers simulations for 
	different energies; dashed lines denote the corresponding Gibbs distribution.}
	\label{fig:pdfThermalisedVelocity}
\end{figure}

Such a prescription leads to the inviscid, three-dimensional (3D) Euler and
one-dimensional (1D) Burgers equations, but retaining only a finite number of
Fourier modes through a (Fourier) Galerkin
truncation~\cite{cichowlas_euler,brachet_two_fluid,ssray_burgers,majda_burgers}.
Such a projection of the partial differential equations on to a
finite-dimensional sub-space ensures conservation of momentum, energy and phase
space, \textit{and} guarantees chaotic solutions for the flow field which
thermalise.  These thermalised fluids (see Appendix A)
are characterised by energy equipartition and velocity fields with Gibbs
distribution $\mathcal{P}[{\bf v}]~d\mathbf{v} = \left( \frac{3 }{2\pi E }
\right)^{3/2} \exp[-{3 \mathbf{v}^2}/2E]~d{\bf v}$ as illustrated in
Fig.~\ref{fig:pdfThermalisedVelocity}. Here $E$ is the conserved energy density
of the system satisfying $\left\langle \frac{1}{2} \mathbf{v}^2 \right\rangle
=E$. This allows us to define a {\it temperature}, $T=\frac{2}{3}E$ such that
the different thermalised configurations describe a canonical ensemble.  A
thermalised fluid is thus not dissimilar to that of correlated many-body
condensed matter systems (e.g., frustrated magnets) where the microscopic
memory does not dictate the dynamical correlations.  

These thermalised fluids set the platform for addressing the primary question
of the growth of perturbations in a classical, chaotic system. To do this, in
the 3D Euler,  an arbitrary realisation of the thermalised solution ${\bf
v_0^a} = {\bf v}^{\rm th}$ is taken and a second copy generated, with a
perturbation in velocity field, ${\bf v_0^b} = {\bf v_0^a} +\bm{ \delta }{\bf
v}_0$. Here, $\bm{ \delta } {\bf v}_0  = \bm{\nabla }\times \mathbf{A}$, with
$A_i= \epsilon \sqrt{E} \; r_{0}\exp \left[ {-\tfrac{r^2}{2r_{0}^{2}}} \right]
\hat{e}_i$, is an infinitesimal (characterised by $\epsilon \ll 1$)
perturbation centred at the origin and which falls off rapidly with distance
$r$ (with the reference scale $r_0\ll 2\pi$) making it \textit{spatially
localised}.

We now evolve (see Appendix B) the Galerkin-truncated Euler equation,
\textit{independently} for the two copies, with initial conditions ${\bf
v_0^a}$ and ${\bf v_0^b}$ to obtain (thermalised) solutions ${\bf v^a}({\bf
x},t)$ and ${\bf v^b}({\bf x},t)$ and thence the \textit{difference field}
$\bm{ \delta } {\bf v}({\bf x},t)  = {\bf v^b}({\bf x},t) - {\bf v^a}({\bf
x},t)$. Since initially this difference field $\bm{ \delta } {\bf v}({\bf x},0)
\equiv \bm{ \delta }{\bf v}_0$ was spatially localised and vanishingly small,
its subsequent spatio-temporal evolution reflects how the butterfly effect
manifests itself in such systems.  Fundamentally, this is  a question of how
systems ${\bf a}$ and ${\bf b}$ decorrelate and intimately connected with
questions of ergodicity and thermalisation.

\begin{figure}
	\centering
	\includegraphics[scale=1.0]{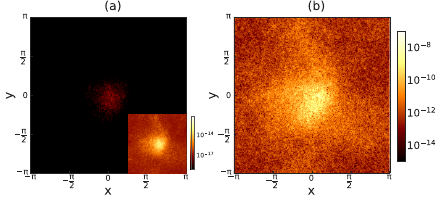}
	\caption{Representative plots of the difference
field $|\bm{\delta }\mathbf{v}(\mathbf{x},t)|^2$, along the
$z=0$ plane of a 3D thermalised fluid (with energy $E = 2.0$ and a perturbation amplitude $\epsilon = 10^{-6}$ 
at (a) early ($t=0.4$) and (b) later ($0.7$) times. 
The inset of panel (a) shows the same early time data, with a magnified scale, 
	to reveal a somewhat {\it self-similar} spatial structure that arises from non-local interactions (see main text). See also \url{https://www.youtube.com/watch?v=yRmdvwX5zhE} for a time evolution of the difference field.}
	\label{fig:sectionOfDecorrelation}
\end{figure}

To make this assessment rigorous, we construct the spatially-resolved
decorrelator   $\phi({\bf x},t)=\left\langle \frac{1}{2} |\bm{ \delta
}{\bf v}({\bf x},t)|^2 \right\rangle$, where $\langle \cdots \rangle$ denotes
averaging over configurations taken from the thermalised ensemble and 
distance is measured from origin where the perturbation is seeded at $t=0$. In
Fig.~\ref{fig:sectionOfDecorrelation} (see, \url{https://www.youtube.com/watch?v=yRmdvwX5zhE} for a video of the full evolution) we show the spatial profile (in
the $z=0$ plane) of $|\bm{\delta} {\bf v}({\bf x},t)|^2$ for a particular
initial realisation of systems {\bf a} and {\bf b} at two different instants of
time.  While at very early times $t= 0 ^+$, panel (a), $|\bm{\delta} {\bf
v}({\bf x},t)|^2$  remains small but diffuses \textit{instantly} and
\textit{arbitrarily}, a more striking behaviour is seen at later times (panel
(b)) when the spatial spread is controlled by the strain in the velocity field
as we shall see below. (It is likely that that the initial,
instantaneous spread is a result of the non-locality (in space) of the 3D fluid
because of the pressure term; however since the Galerkin-truncation also
introduces an additional non-locality, the precise mechanism for the initial
spread is hard to pin down.)

Since the thermalised fluid is \textit{statistically} isotropic, the
decorrelator $\phi(\mathbf{x},t)$ is a function of $|\mathbf{x}|$.  We exploit
this to construct the more tractable angular-averaged decorrelator
$\phi(r,t)=\frac{1}{4\pi r^2}\int d\boldsymbol{\Omega}_r~\phi({\bf x},t)$.  

Given the non-locality of the 3D Euler equation, these systems differ crucially
from spin systems in the absence of pilot waves and a distinct velocity scale
akin to a butterfly speed~\cite{Das2018,PhysRevLett.121.250602}. Instead,
decorrelators for 3D thermalised fluids have a self-similar
spatial profile $\phi(r,t)\sim r^{-\alpha }$ (with $\alpha\sim 4$). The lack of
a sharp wave-front and self-similarity is evident from Fig.
\ref{fig:sectionOfDecorrelation}(b) and the inset of Fig.
\ref{fig:sectionOfDecorrelation}(a). Therefore to track the temporal evolution
of the decorrelator it is convenient to introduce the space-averaged
decorrelator $\Phi(t)=\frac{1}{V} \int d^3{\bf x}~\phi({\bf x},t)$ which then
serve as a diagnostic for the temporal aspects of this problem.

This allows us, starting from the 3D Euler equation, to derive the evolution equation
\begin{align}
\dot{\Phi}(t)=-\overline{\left\langle \delta v_i S_{ij} \delta v_j \right\rangle}
\label{eq:strain-decor}
\end{align}
in terms of the familiar rate-of-strain tensor $2S_{ij} = \partial_i v^{\rm a}_j + \partial_j v^{\rm a}_i$; 
the over-bar in the definition denotes a spatial average.

By using the eigenbasis of ${\bf S}$, we re-write the above equation as 
$\dot{\Phi}=- \displaystyle \sum_{i=1}^{3}\:\overline{\left\langle \hat{\alpha }_i^2 \gamma _i |\bm{\delta
}\mathbf{v}|^2 \right\rangle} $ where $\left\{ \gamma_i  \right\}$ are the
three eigenvalues and $\left\{ \hat{\alpha_i } \right\}$ are the direction
cosines of $\bm{\delta }\mathbf{v}$ along the three eigen-directions.
Equation~\ref{eq:strain-decor}, which formally resembles the enstrophy
production term for the Euler equation~\cite{ashurst_87,gibbon_97}, is an
important result that \textit{connects} the decorrelator with the dynamical
structures of the velocity field.

Our direct numerical simulations (DNSs, see Appendix B) of the truncated 3D Euler
equation show strong evidence that the difference fields preferentially grow,
at \textit{short} times, along the compressional eigen-direction ($i = 3$) of
the thermalised fluid leading to a further simplification $\dot{\Phi} \approx
-\overline{\left\langle \hat{\alpha}_{3}^{2}\gamma _{3}|\bm{\delta
}\mathbf{v}|^2 \right\rangle}$.  Since by definition $\gamma_3 < 0$, this
ensures not only the positive definiteness of $\dot{\Phi}(t)$, but also, since
(up to constants) $\dot{\Phi}(t) \sim -\overline{\gamma _{3}} ~\Phi$, an
exponential growth with a Lyapunov exponent $\lambda\sim |\overline{\gamma_3}|$
at short times (Fig.~\ref{fig:autocorrelationFunction}). This connects the
straining of the flow-field with $\lambda$.

How robust is this \textit{short-time} behaviour with respect to both dimension and the compressibility of the flow?

The answer lies in an analysis of the 1D (compressible) Burgers
equation with $N_G$ Fourier modes.  Furthermore, to underline the universality
of our results, this time we construct the decorrelator and carry out the
theoretical (see Appendix C) and numerical analysis entirely in Fourier space. As before, from the
thermalised solution (in Fourier space) ${\hat v}^{\rm th}$, defining a control
field ${\hat v}^{\bf a}_0 = {\hat v}^{\rm th}$ and a perturbed field ${\hat
v}_0^{\bf b} = {\hat v}_0^{\bf a}(1 + \epsilon\delta_{k,k_p})$ with large
values of the perturbation wave-number $k_p$ to generate de-localised
small-scale perturbations in the systems (see Appendix B).  It is
important to stress that given the seed perturbation is localised in Fourier
space in 1D (and hence de-localised in physical space), the spatial spread of
perturbations, which is relevant and studied for 3D fluids in this Letter,
remains outside the scope of analysis here.

As before, both systems are evolved independently and the Fourier space
decorrelator $|\hat\Delta_k|^2 =  \langle|\hat{v}_k^{\bf a} - {\hat v}^{\bf
b}_k|^2\rangle$, measured, mode-by-mode, as a function of time.  Given the
relative analytical simplicity of the 1D system, we construct the equation of
motion of $|\hat\Delta_k|^2$  and derive an exponential growth of the
decorrelator associated with a Lyapunov exponent $\lambda$. Thus the
theoretical calculations for the 1D model are not only consistent with the more
complex 3D system but also provide, as we see below, a more rigorous insight
into how the Lypunov exponent scales with  $T$ and the degrees of freedom $N_G$
of our system. (See Appendices C -- F for the derivation of
the linear theories describing the short-time dynamics of the decorrelator.)

\begin{figure} \centering
	\includegraphics[scale=0.8]{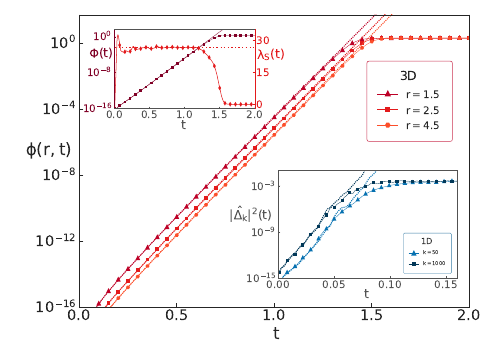} 
	\caption{Semi-log plots of $\phi (r,t)$ ($E = 1.0$)) and (lower inset) $|\hat\Delta_k|^2$ ($E = 2.0$) showing an initial exponential 
	growth and eventual saturation. The dashed lines, from linearised theory, 
	are in excellent agreement with DNSs at early times.  
	(Upper inset) Semi-log (left y-axis) plot of $\Phi (t)$ (3D fluid) along with results from our
	linearised theory (dashed line). $\lambda$, extracted from  $\Phi(t)$, shown as dash-dot horizontal line, agrees well 
	with $\lambda_S$ (linearised theory, right y-axis).}
	\label{fig:growthOfDecorrelator}
\end{figure}

At long times, since systems {\bf a} and {\bf b} decorrelate $\langle {\bf
v^a}\cdot{\bf v^b} \rangle = 0$, leading to suspension of the underlying
approximations in the linear theory presented above, $\Phi$ and
$|\hat\Delta_k|^2$ must saturate to a value equal to $2E$ and $2E/N_G$
respectively. 

With these theoretical insights for both the 1D and 3D systems, we
test them against results from our numerical simulations. In
Fig.~\ref{fig:growthOfDecorrelator} we show representative results for 
$\phi(r,t)$ ($\Phi$ in the upper inset) from 3D Euler and $|\hat\Delta_k|^2$ for the 
1D Burgers (lower inset) versus time on a semi-log scale. The symbols (for
different values of $r$ and $k$) are results from the full nonlinear DNSs while
the dashed lines correspond to decorrelators obtained the linearised theory.

Consistent with our theoretical estimates described above, the decorrelators
from the full, nonlinear DNSs (shown by symbols) grow exponentially (positive
$\lambda$) before eventually saturating (as the two systems decorrelate) to a
value set by the energy.  The agreement between these decorrelators and the
ones we estimate theoretically through a linearised model (dashed lines) is
remarkable during the early-time exponential phase.  However, decorrelators
constructed from the linearised model (valid for short times) are insensitive
to  non-linearities and continue growing exponentially, while the ones from the
full nonlinear system eventually saturate. We will soon return to the question
of time scales which determine this saturation.

Finally, we confirm the  validity of Eq.~\ref{eq:strain-decor} by showing
(upper inset, Fig.~\ref{fig:growthOfDecorrelator}) the agreement between
$\lambda_S(t) = {-\overline{\left\langle \delta v_i S_{ij} \delta v_j
\right\rangle}}/{\Phi(t)}$ and the Lyapunov exponent $\lambda$  extracted from
the decorrelator $\Phi (t)$ measured in DNSs. The agreement between the two is
almost perfect at short times before $\lambda_S(t)$ decays to zero as the
decorrelator saturates.

This inevitably leads us to central question of this work: How fast do
perturbations grow in a classical, chaotic system and how does it depend on the
temperature $T$ as well as the number of modes, $N_G$ ? Furthermore is the
scaling behaviour of $\lambda$ really universal? 

Although non-linear equations for hydrodynamics do not yield easily to an
analytical treatment, it is tempting to theoretically estimate the functional
dependence of $\lambda$ on $T$ and $N_G$.  An extensive analysis (see Appendices C -- D) of the
linearised equations for $\Phi(t)$ and $|\hat\Delta_k|^2$ show that under very
reasonable approximations,  which were tested against data, $\lambda \propto
N_G\sqrt{T}$.  Whereas for the Euler fluid this scaling is a consequence of the
statistics of the strain-rate-tensor which determines the behaviour of
$\Phi(t)$, the analogous result for the 1D system is obtained by
straightforward algebraic manipulations, factoring in the statistical
fluctuations, of the equation governing the evolution of $|\hat\Delta_k|^2$.

\begin{figure} \flushleft 
\includegraphics[scale=0.8]{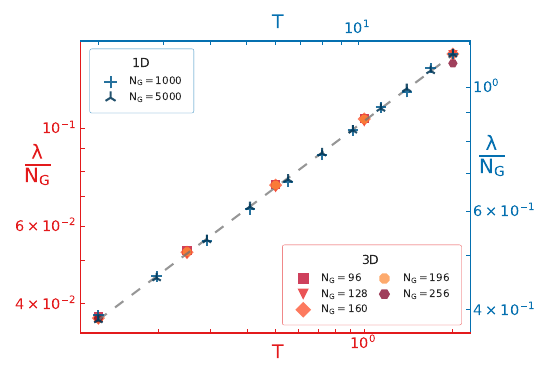}
\caption{Log-log plot of $\lambda/N_G$ versus $T$ for the 3D
(axes in red) and 1D (axes in blue) thermalised fluids corresponding to different values of $\epsilon$, $N_G$ and, 
	for the 1D fluid, $k_p$. The dashed line $\propto \sqrt{T}$ confirms our theoretical 
prediction.} 
\label{fig:lyapunovOfDecorrelator} 
\end{figure}

\begin{figure}
	\centering
	\includegraphics[scale=0.8]{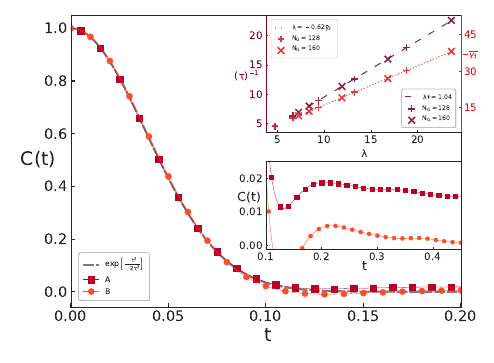}
	\caption{A plot of $C(t)$ for a  
(A) \textit{nearly} and (B) \textit{completely} thermalised 3D fluid along with the theoretical Gaussian prediction.
	(Lower inset) A magnified view shows that for a fluid which is not completely 
thermalised, $C(t)$ falls off to zero much more slowly. 
	(Upper inset) Representative plots of $\tau$ and the average (negative) eigen-value (compressional direction) versus $\lambda$.}
	\label{fig:autocorrelationFunction}
\end{figure}

Our theoretical prediction is easily tested by measuring $\lambda$ in DNSs of
the full non-linear 3D Euler and 1D Burgers equations.  From plots such as in
Fig.~\ref{fig:growthOfDecorrelator}, we extract the mean $\lambda$ and its
(statistical) error-bar, and examine its dependence on temperature $T$ (and
$N_G$) by changing the magnitude of the initial conditions and hence the
initial energy or temperature.  (Surprisingly, $\lambda$ measured through such
decorrelators are independent of $r$ or $k$, as was already suggested in
Fig.~\ref{fig:growthOfDecorrelator}.) Figure~\ref{fig:lyapunovOfDecorrelator}
shows a unified (3D Euler and 1D Burgers) log-log plot of all the rescaled
Lyapunov exponent $\lambda/N_G$ measured---for different strengths and scales
of perturbations and $N_G$---as a function of temperature $T$. The collapse of
the data on the dashed line, denoting a $\sqrt{T}$ scaling, shows that the
many-body chaos of such thermalised fluids \textit{is} characterised by the
behaviour $\lambda \propto N_G\sqrt{T}$. It is worth stressing
that these DNS results for the 3D Euler equations make the theoretical
bound (see Appendix D) sharp.

These, to the best of our knowledge, are the first results, and confirmation of
earlier conjectures~\cite{Kurchan2016,Maldacena2016} and demonstrations for
classical spin systems \cite{PhysRevLett.121.250602}, that $\lambda \propto
\sqrt{T}$ in a chaotic and non-linear, many-body classical system  obeying the
equations of hydrodynamics.  Remarkably, we also find strong evidence that
$\lambda$ scales linearly with $N_G$ in such extended systems and
independent of spatial dimension and compressibility of the flow. 

Given the association of many-body chaos with ergodicity and equilibration in
classical statistical physics, how well do measurements of $\lambda$ relate to
the (inverse) time scales associated with the loss of \textit{memory}?  The
simplest measure of how fast a system \textit{forgets} is the ensemble-averaged
autocorrelation function $C(t)= \left( 2E \right)^{-1}\left\langle
\mathbf{v}^{\mathrm{th}}(t)\cdot \mathbf{v}^{\mathrm{th}}(0) \right\rangle$
(Fig.~\ref{fig:autocorrelationFunction}).  It is easy to show (see Appendices C -- D) that
$C(t) \approxeq \mathrm{exp}\left[-\frac{t^2}{2\tau ^2}\right]$ with an
auto-correlation time $\tau \sim 1/\lambda$ as clearly shown from our
measurements (upper inset,  Fig.~\ref{fig:autocorrelationFunction}). This
association of $\tau$ with $\lambda$ provides a firm foundation to
interpret the salient features of many-body chaos in terms of principles of
statistical physics: Ergodicity and thermalisation.  A further connection is
established through the relation between the time scales $t_{\mathrm{sat}}\sim
\tau \sim 1/\lambda$ at which the decorrelator saturates as $\Phi(t)\approxeq
\;2E \left( 1+\mathrm{exp}\left[-\lambda (t-t_{\mathrm{sat}})\right]
\right)^{-1}$.

The generality of the OTOCs and cross-correlators lead to questions of
connecting the macroscopic variables with the scales of chaos in the most
canonical of chaotic systems: Those described by non-linear equations of
hydrodynamics. Here we provide the first evidence of the temperature dependence
of the Lyapunov exponent in (continuum) classical non-linear hydrodynamic
systems and show its robustness with respect to spatial dimensions and
compressibility effects. It is important to underline that many-body chaos and
$\lambda \sim N_G \sqrt{T}$ is really an emergent feature of a fluid which
\textit{is} thermalised. We checked this explicitly by measuring the
decorrelators in the flow \textit{before} it thermalises and found, despite the
conservation laws still holding, no associated exponential growth and spread of
the difference field.  Furthermore, our measured $\lambda$ should be identified
with the largest Lyapunov exponent of the system and that $t_{\mathrm{sat}}$ is
a useful estimate of thermalisation (or equilibration) time. 

Finally, the temperature dependence of $\lambda$
is consistent with recent results for classical spin liquids without
quasi-particles~\cite{PhysRevLett.121.250602,Scaffidi2017,ruidas2020many,bilitewski2020classical} as well as more
general dimensional arguments based on phase-space dynamics~\cite{Kurchan2016} 
of classical many-body systems. In this regard we note that in classical
spin-systems~\cite{bilitewski2020classical}, the existence of low energy quasi-particles seems to {\it reduce} the
chaotic behaviour of the system ($\lambda \propto T^a,~a>0.5$). While more
detailed and theoretical investigation of these features, as well as, how far
they are relevant for the spontaneously stochastic Navier-Stokes
turbulence are interesting future directions, this 
\textit{butterfly effect} for classical, non-linear, hydrodynamic systems seems
to be robust and generic. 

While it is probably true that the exact nature of the dependence of the
Lyapunov exponent on the temperature (or energy density) and number of
degrees of freedom should vary from system to system, the evidence we
provide of their inter-dependence opens new avenues and questions.
In particular, these studies demonstrate the dependence of signatures of
spatio-temporal chaos on the thermodynamic variables as well its relation with
the transport properties. 

\begin{acknowledgements}

We thank S. Banerjee, J. Bec, M. E. Brachet, A. Dhar, A. Kundu, A.
Das, S. Chakraborty, T. Bilitewski, R. Moessner, and V. Shukla for insightful
discussions. The simulations were
performed on the ICTS clusters {\it Mowgli}, {\it Mario}, {\it Tetris}, and
{\it Contra} as well as the work stations from the project ECR/2015/000361:
{\it Goopy} and {\it Bagha}. DK and SSR acknowledges DST (India) project DST
(India) project MTR/2019/001553 for financial support. SB acknowledges MPG for
funding through the Max Planck Partner group on strongly correlated systems at
ICTS. DK and SB acknowledges SERB-DST (India) for funding through project grant
No. ECR/2017/000504. This research was supported in part by the International Centre for Theoretical Sciences (ICTS) for the online program - Turbulence: Problems at the Interface of Mathematics and Physics (code: ICTS/TPIMP2020/12). The authors acknowledges the support of the DAE, Govt. of India, under project no.  12-R\&D-TFR-5.10-1100 and project no. RTI4001. 

\end{acknowledgements}

\appendix

\section{Appendix A: Thermalised fluids: Galerkin projection of 1D Burgers and 3D Euler Equations}

The dynamics of inviscid, ideal fluids satisfy well-known partial differential equations. For 
the scalar velocity field $u$ in one-dimensional (1D) flows, this is known as the (inviscid) Burgers 
equation:
\begin{equation} \frac{\partial u}{\partial t} + \frac{\partial u^2}{\partial
x} = 0 \label{burgers} \end{equation}
with initial conditions  $u_0$.

For three-dimensional flows (3D), the analogous equation for the (incompressible) velocity vector ${\bf u}({\bf x},t)$ and scalar
pressure $P$ fields satisfy the celebrated Euler equation:
\begin{equation} \frac{\partial {\bf u}}{\partial t} + {\bf u}.\cdot \nabla
{\bf u} = -\nabla P \label{euler} \end{equation}
augmented by the constraint $\nabla\cdot{\bf u} = 0$ and initial conditions
${\bf u}_0$. 

While the 1D inviscid Burgers equation admits real singularities, which
manifests itself as pre-shocks and then shocks in the velocity profile at a
finite time $t_\star$ (Fig.~\ref{fig:SI}(a)) and dissipates energy (even in the absence of a viscous
term)~\cite{frisch2001houches,burg-review-bec}, the issue of finite-time
blow-up for the 3D Euler equation still remains one of \textit{the} most
important, \textit{unsolved}, problems in the natural sciences.  Nevertheless,
weak (in the sense of distributions) solutions of the 3D Euler equations have
been recently shown to be dissipative as conjectured by Onsager and consistent
with the celebrated problem of \textit{dissipative anomaly} of high Reynolds
number turbulence. Therefore, such inviscid, infinite-dimensional partial
differential equations, in one or three dimensions (like their viscous
counterparts) lack a Hamiltonian structure and cannot lead to solutions
characterised by a statistical equilibria.

Fortunately, a subtle, but significant, modification to
these equations, while preserving the essential nonlinearity, allows us to move
away from the dissipative to thermalised solutions with an energy equipartition
and Gibbs distribution of the velocity field.  Within the space of $2\pi$
periodic solutions, an expansion of the solution in an infinite Fourier series
allows us to define the Galerkin projection as a low-pass filter
$\mathrm{P}_{K_{G}}$ which sets all modes with wave vectors $|{\bf k}| >
K_{\scriptscriptstyle \mathrm{G}}$, where $K_{\scriptscriptstyle \mathrm{G}}$
is a positive (large) integer, to zero via $\mathrm{P}_{K_{G}} {\bf u}({\bf x})
=   \sum_{|{\bf k}| \le K_{\scriptscriptstyle \mathrm{G}}} \mathrm{e}
^{\mathrm{i} {\bf k}\cdot{\bf x}}\hat {\bf u_k}.$ The truncation wavenumber
$K_{\scriptscriptstyle \mathrm{G}}$ sets the number of Fourier modes $N_G$ kept
and is a measure of the effective number of degree of freedom as well as
providing a microscopic (ultraviolet) cut-off for the system. These
definitions, without the loss of incompressibility, lead to the
Galerkin-truncated Euler equation for the truncated field, written, most
conveniently, component-wise in Fourier space ${\hat v}_\alpha({\bf k})$
\begin{equation} \frac{\partial {\hat v}_\alpha({\bf k})}{\partial t} =
-\frac{\mathrm{i}}{2}\mathcal{P}_{\alpha\beta\gamma}({\bf k})\sum_{{\bf
p}}{\hat v}_\beta ({\bf p}){\hat v}_\gamma ({\bf k} - {\bf p}); \label{gteuler}
\end{equation}	
where the initial conditions ${\bf v}_0 = \mathrm{P}_{K_{G}} {\bf u}_0$ and the
convolution $|{\bf k} \le K_{\scriptscriptstyle \mathrm{G}}$, $|{\bf p}| \le
K_{\scriptscriptstyle \mathrm{G}}$ and $|{\bf k}-{\bf p}| \le
K_{\scriptscriptstyle \mathrm{G}}$ is constrained via Galerkin truncation.  The
coefficient  ${\mathcal P}_{\alpha\beta\gamma}  = k_\beta P_{\alpha\gamma} +
k_\gamma P_{\alpha\beta}$, where $P_{\alpha\beta} = \delta_{\alpha\beta} -
k_\alpha k_\beta/k^2$ factors in the contribution from the pressure gradient
and enforces incompressibility; $\delta_{\alpha\beta}$ is a Kronecker delta.

The same definitions of Galerkin truncation can be extended {\textit mutatis
mutandis} to one dimension, without the additional constraints of
incompressibility or pressure gradients, to similarly project the 1D inviscid,
$2\pi$-periodic Burgers equation onto the subspace spanned by
$K_{\scriptscriptstyle \mathrm{G}}$:
\begin{equation} \frac{\partial {\hat v}(k)}{\partial t} = -\frac{\mathrm{i}
k}{2}\sum_{p} {\hat v}(p){\hat v}(p-k).  \label{gtburgers} \end{equation}
With initial conditions $v_0$, the Galerkin-truncated Burgers equation also
imposes the constraint $k \le K_{\scriptscriptstyle \mathrm{G}} $, $p\le
K_{\scriptscriptstyle \mathrm{G}}$ and $|p-k| \le K_{\scriptscriptstyle
\mathrm{G}}$ on the convolution.  

Thus, beginning with the partial differential equations of ideal fluids in one
and three dimensions, Galerkin truncation leads, by self-consistently
restricting the velocity field to a finite number of modes $N_G$, to a
finite-dimensional, nonlinear dynamical systems with a mathematical, nonlinear
structure identical to the equations which govern turbulent flows. However,
such a truncation, which conserves phase space volume,  momentum and kinetic
energy, results, through Liouville's theorem, in solutions at finite times
which are in statistical equilibria (unlike the non-equilibrium steady states
associated with turbulence) with a characteristic Gibbs distribution and a
broadening (standard deviation) $\sigma$ determined by the total (conserved)
energy $E$ (Fig. 1, main paper).  Thence, a natural association of a
temperature $T$ via $T = 2\sigma^2  = 2E/3$ for such systems.  

Furthermore, it is because of this statistical equilibria that
such solutions show an equipartition of kinetic energy amongst all its Fourier
modes resulting in, for the 3D problem, an (shell averaged) energy spectrum
$E(k) = | {\hat v}^{\rm th}({\bf k})|^2 \propto k^2$~\cite{cichowlas_euler}
(or, in 1D, $E(k) = |{\hat v}^{\rm th}_k|^2 \propto k^0$~\cite{ssray_burgers,Ray-Review,ssray_divya,Murugan})
at odds with the well known $k^{-5/3}$ spectrum of real turbulence (or $k^{-2}$
in solutions of the Burgers equation in the limit of vanishing viscosity) as clearly 
shown in Fig.~\ref{fig:SI}(a). 

Thus these chaotic systems (in one or three dimensions), rooted in the
nonlinear equations of hydrodynamics which form the basis of real turbulence
and yet remain in statistical equilibria provides an excellent model for a
thermalised fluid. Furthermore, given the conservation of energy and its
association with temperature through the Gibbs distribution, it is simple to
generate thermalised flows with different temperatures by a simple change in
the amplitude of the initial conditions.

\section{Appendix B: Direct Numerical Simulations (DNS) of truncated equations}
We perform direct numerical simulations (DNSs) of these Galerkin-truncated 3D Euler and the 1D Burgers
equations by using a standard pseudo-spectral method with a
fourth order Runga-kutta algorithm for time marching.  These equations are
solved on a $2\pi$ periodic domain with $N^3$ for the 3D and $N$ for the 1D equations 
with a truncation wavenumber $K_G$ which results in ${N_G}^3 < N^3$ 
(or, in 1D, $N_G < N$) number of degrees of freedom. In our numerical simulations, 
we have explicitly checked that the kinetic energy is conserved and within a finite 
time energy equipartition is reached.

For the 3D truncated Euler problem, we begin with an initial kinetic energy
spectrum of the form $E(k) =  A_0k^4 \exp \left[ -\tfrac{k^2}{2k_{0}^{2}}
\right]$; changing the numerical value of the factor $A_0$ allows us to
generate thermalised fluid with different energies $E$ and hence temperatures $T$. In our simulations,
we use different resolutions $N^3 = 96^3, 128^3, 160^3, 196^3 , 224^3$ and
$256^3$ for different values of the truncation wavenumbers
$K_{\scriptscriptstyle \mathrm{G}} = N/3, N/4$ and $N/5$  to generate flows
with varying degrees of freedom $N_G\sim K_{\scriptscriptstyle \mathrm{G}}$ as well as 
different amplitudes of the initial conditions to scan the temperatures in the range $0.125 \leq T \leq 4$, .
Our time-step for integration, depending on $K_{\scriptscriptstyle \mathrm{G}}$
and $E$, varies as $ \Delta t \lesssim \sqrt{\frac{3}{2E}}\frac{2\pi }{N}$, and
the truncated equations were integrated up to a time $t \approx 10$ to generate
fully thermalised solution ${\bf v}^{\rm th}$ which provides the starting point
to generate systems ${\bf a}$ and ${\bf b}$ used in our calculations of the
decorrelator $\Phi(t)$. 

For the 1D truncated Burgers problem, we choose an
initial condition $v_0 = A_0[\sin x + \sin (2x-0.2) + \sin (5x-0.4) + \sin
(7x-0.5)]$; the precise functional form of the initial conditions is immaterial
with the total conserved momentum $\int_0^{2\pi} v_0~dx=0$. Further (as in the 3D problem), changing
the numerical constant $A_0$, allows us to change the energy $E =
\frac{1}{2\pi}\int_0^{2\pi} v_0^2dx = 2A_0^2$ of our system and thence the
temperature $2 \leq T \leq 18$.  Given the lower computational cost for solving the 1D system,
we were able use a much larger number of collocation points $N = 2^{14}$ to
generate systems with larger values of $K_{\scriptscriptstyle \mathrm{G}} =
1000$ ($\delta t = 10^{-5}$) and 5000 ($\delta t = 10^{-6}$) leading to values
of $N_G$ much larger than those accessible to 3D simulations.

To perturb the system ${\bf a}$, we introduce, for the 3D fluid, a
perturbation of strength $\epsilon = 10^{-6}$; in the 1D problem, we use
$\epsilon = 10^{-5}$ and $10^{-4}$. Furthermore, since the perturbation, for
the 1D problem, is introduced at wavenumber $k_p$ in the Fourier space, we
choose different values of $k_p = 500, 900, 2500$ and $4000$ to demonstrate the
insensitivity of our results to the precise (small) scales of perturbation (and $\epsilon$) 
as clearly seen in the collapse of the data in Fig.~3 of the main text.\\

\section{Appendix C: Decorrelators: The Linearised Theory}

Systems {\bf a} and {\bf b} both satisfy the Galerkin-truncated, three-dimensional (3D) Euler equation. Therefore 
the evolution equation for the \textit{difference field} 
 $\delta {\bf v}({\bf x},t) = {\bf v^b}({\bf x},t) - {\bf v^a}({\bf x},t)$, component-wise 
is given by: 
 \begin{eqnarray} \partial_{t}\delta v_i\left( \mathbf{x}, t \right) &=&
	 -\partial_{j}\left[ v_{i}^{\bf a}\delta v_j + v_j^{\bf a}\delta v_i +
	 \delta v_i \delta v_j \right] \nonumber \\ &+& \partial _{ijl}^{3}\displaystyle
	 \int_{\mathcal{D}}^{}d\mathbf{x}^{\prime}\:
	 G(\mathbf{\mathbf{x},\mathbf{x}^{\prime}})\left[ v_{j}^{\bf a}\delta
	 v_l+ v_{l}^{\bf a}\delta v_j+\delta v_j\delta v_l \right]^{\prime}; \nonumber \\
 \label{evolutionPerturbation} \end{eqnarray}
with an initial conditions $\bm{ \delta } {\bf v}({\bf x},0) \equiv \bm{ \delta }{\bf v}_0$
and a Green's function satisfying 
$\bm{\nabla }^2
 G(\mathbf{x},\mathbf{x}^{\prime})=\bm{\delta}(\mathbf{x}-\mathbf{x}^{\prime})$. 
While the non-local and convective terms in this equation clearly suggests that
a localised, initial difference $\bm{ \delta } {\bf v}({\bf x},0) \equiv \bm{ \delta }{\bf v}_0$, introduced 
through the perturbation in {\bf b}: ${\bf v_0^b} = {\bf
v_0^a} +\bm{ \delta }{\bf v}_0$; with $\bm{ \delta } {\bf v}_0  = \bm{\nabla
}\times \mathbf{A}$, where $A_i= \epsilon \sqrt{E} \; r_{0}\exp \left[
        {-\tfrac{r^2}{2r_{0}^{2}}} \right] \hat{e}_i$, will de-localise with a spatio-temporal 
spreading. However, given the nonlinear nature of this equation, estimating how this happens, or more 
specifically, the temporal growth of the decorrelator $\Phi(t)$ and thence the Lyapunov exponent, is 
a challenge. 

Since the main question which concerns us has to do with the \textit{short}
time growth of these decorrelators, when nonlinear terms $\mathcal{O}(\delta v^2)$
can be ignored, a reasonable assumption which was validated against data from our Direct Numerical Simulations (DNSs), 
we \textit{linearise} Eq.~\ref{evolutionPerturbation}:

\begin{equation} \frac{\partial \delta v_i^{\mathrm{lin}}}{\partial t} \approx
	-v_j^{\bf a}\frac{\partial \delta v_i^{\mathrm{lin}}}{\partial x_j} -
	\delta v_j^{\mathrm{lin}}\frac{\partial v^{\bf a}_i}{\partial x_j} +
	\frac{\partial T}{\partial x_i}; \label{linearisedEuler} \end{equation}
where, $T = 2\delta_{ij}^2\int d{\bf y}G(|{\bf x}-{\bf y}|)\delta
v_j^{\mathrm{lin}}({\bf y})v_k^{(a)}({\bf x})$ is the non-local (linear)
contribution from the pressure term. It is worth stressing that although we
linearise the equation, it still allows for the spatio-temporal spread of the
difference field because of its non-local nature. As we have shown in the main
paper (Fig. 3; dashed lines), the decorrelator $\Phi(t)$ (or $\phi(r,t)$)
obtained from this linearised equation is in agreement, at short times, with
those obtained from the DNSs of the full 3D truncated Euler equation. Indeed,
quantifying by this \textit{agreement} through a global relative error:
\begin{equation} \Gamma (t)= \dfrac{\frac{1}{V}\displaystyle
\int_{\mathcal{D}}^{}d\mathbf{x}\frac{1}{2} |\bm{\delta }\mathbf{v}-\bm{\delta
}\mathbf{v}^{\mathrm{lin}}|^2\: }{\Phi(t)} \label{errorDef} \end{equation}
where $\mathcal{D}$ is the domain and $V$ the volume of space; in the exponential
growth regime $\Gamma \sim 10^{-4}$ and reaches $\mathcal{O}(1)$ at times when
the $\Phi(t)$ (or $\phi(r,t)$) obtained from DNSs start to saturate. The linear
theory of course fails in this saturation region as the approximations leading
upto it no longer holds as $\langle {\bf v^a}\cdot{\bf v^b} \rangle = 0$ and
$|\bm{\delta }\mathbf{v}|$ is of the same order as the root-main-squared velocity
of the thermalised fluid. 

Nevertheless, starting with Eq.~\ref{evolutionPerturbation} and taking dot
products with $\delta {\bf v}({\bf x},t)$ followed by a spatial integration, we
eventually obtain: 
\begin{equation} 
\dot{\Phi}(t) = -\left\langle \delta v_i S_{ij} \delta v_j
\right\rangle + \left\langle \partial_{j} W_j \right\rangle
\label{strainCouplingP1} 
\end{equation} 
with 
\begin{equation} W_{j} =
	-\frac{1}{2}v_{j}^{\bf b}|\bm{\delta }\mathbf{v}|^2+\delta v_j \partial
	_{il}^{2}\displaystyle \int_{\mathcal{D}}^{}d\mathbf{x}^{\prime}\: (
	v_i^{\bf a}\delta v_l + \delta v_i v_{l}^{\bf b}
	)^{\prime}G_{\mathbf{x},\mathbf{x}^{\prime}} \end{equation}
and $S_{ij}$ the familiar rate-of-strain tensor $2S_{ij} = \partial_i v^{\bf
a}_j + \partial_j v^{\bf a}_i$ for the thermalised fluid.
The second, divergence term in Eq.~\ref{strainCouplingP1} vanishes however because of periodic 
boundary conditions leading to 
\begin{align} \dot{\Phi}(t)&= - \left\langle \bm{\delta }\mathbf{v}\cdot
\mathbf{S} \cdot \bm{\delta }\mathbf{v} \right\rangle
\label{strainCouplingP2} \end{align}
In the main paper, we have illustrated the validity of Eq.~\ref{strainCouplingP2} in the upper inset of 
Fig. 5. 

Since the rate-of-strain tensor is diagonalisable, in its eigenbasis with  eigenvalues
$\left\{ \gamma_i \right\}$ (satisfying the incompressibility constraint
$\sum_i \gamma_i = 0$ with extensional $\gamma_1 > 0$ and compressional $\gamma _3<0$ eigen-directions) 
we decompose $\delta {\bf v}$ in the eigenbasis of $S_{ij}$
with (undetermined) components $\alpha_i$ along each eigenvector:
\begin{align} \dot{\Phi}(t)&= -\displaystyle \sum_{i=1}^{3}\: \left\langle
\alpha _{i}^{2}\gamma _i \right\rangle \label{strainCouplingP3} \end{align}
Keeping in mind that the thermalised fluid is incompressible and $\dot{\Phi}(t) > 0$ at short and 
$\dot{\Phi}(t) = 0$ at long times (saturation), $\alpha _i^2$ are clearly correlated with the 
corresponding eigenvectors. Further more, since  $\gamma _3<0$ and $\dot{\Phi}(t)$ is positive at short 
times, it seems likely that there \textit{must} be a preferential
alignment of $\bm{\delta }\mathbf{v}$ with the compressional eigenvector. 

\begin{figure*}
	 	\includegraphics[scale=1.0]{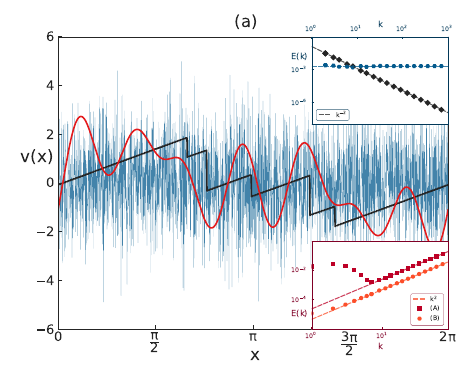}
	 	\includegraphics[scale=1.0]{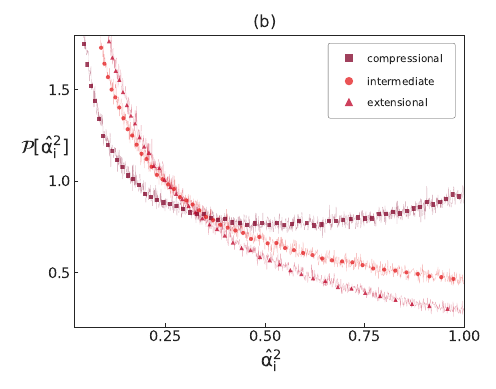}\\
	 	\includegraphics[scale=1.0]{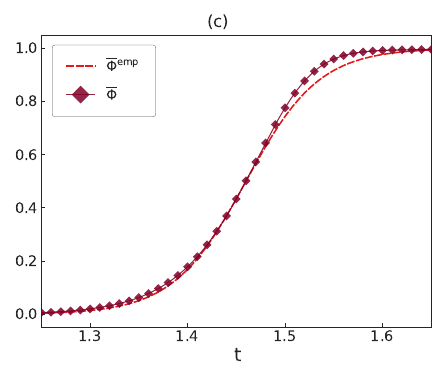}
	 	\includegraphics[scale=1.0]{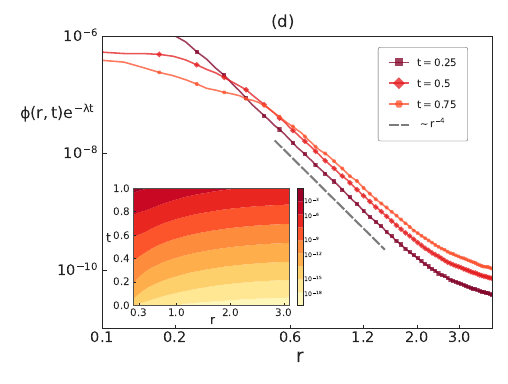}
	\caption{(a) 
	Plots of the Galerkin-truncated Burgers equation velocity field $v$ (blue) and the 
	un-truncated solution $u$ (black) at time $t = 10.0$ 
	for an initial condition (red) with $A_0 = 1.0$. The un-truncated or entropy (weak) solution shows clear 
	shocks and the characteristic saw-tooth profile whereas the truncated solution appears as a \textit{white noise} indicative 
	of a thermalised flow. 
	(Upper Inset) Log-log plots of energy spectrum \textit{vs} $k$  
	for the truncated (blue circles) and un-truncated (black squares) 
	equations showing the equipartition of 
	kinetic energy ($E(k) \sim k^0$) for the former and a $k^{-2}$ scaling (denoted by a thick black line) for the latter. (Lower 
	Inset) The kinetic energy spectrum of a partially (A, squares; early times) and fully (B, circles; late times) thermalised 3D fluid 
	obtained from simulations of the Galerkin-truncated Euler equation. The dashed lines with a scaling $\sim k^2$ is indicative of 
	energy equipartition due to thermalisation. (b) The probability distribution of the components $\hat{\alpha}_i^2$ for the three 
	eigen-directions clearly suggests the preferential alignment along the direction of compression. (c) A representative plot 
	of the decorrelator $\Phi(t)$ for a 3D thermalised flow ($E = 1$) and the empirical form $\overline{\Phi}^{\mathrm{emp}}(t)$ 
	illustrating the approximate agreement between the two. (d) The compensated decorrelator $\phi(r,t)\exp(-\lambda t)$ for different 
	values of $t$ ($E = 1$); the grey dashed line shows a scaling $r^{-4}$ is an illustration of the self-similar nature of the spatial 
	spread of the decorrelations. This is confirmed (inset) in the space-time plot of the isocontours of the decorrelator which 
	suggests  a  spread of the forms $t\sim \ln r$.} 
	\label{fig:SI}
\end{figure*}

Theoretically, this idea of preferential alignment is hard to prove. However,
we are able to construct from our numerical data the probability distributions
of the $\alpha_i$ (Fig.~\ref{fig:SI}(b)) for all three eigen-directions
and find that the conjectured preferential alignment, namely that the sum in the right hand side of
Eq.~\ref{strainCouplingP3} is dominated by $\gamma_3 < 0$ leading to
$\dot{\Phi} > 0$, holds.

This allows us to simplify the equation of motion of the decorrelator
	\begin{align}
		\dot{\Phi}(t) & \sim -\left\langle \alpha _{3}^{2}\gamma _3 \right\rangle \sim - \overline{\gamma}_3  \Phi
		\label{strainCouplingP31}
	\end{align}
with $\overline{\gamma}_3$ the average (negative) eigenvalue along the compressional direction.

For the thermalised fluid emerging from solutions of the Galerkin-truncated 1D Burgers equation, the linearised 
theory for the decorrelator is relatively straightforward.  
We recall that beginning with the thermalised solution (in Fourier space) ${\hat v}^{\rm th}$ allows us to define a
control field ${\hat v}^{\bf a}_0 = {\hat v}^{\rm th}$ and a perturbed field
${\hat v}_0^{\bf b} = {\hat v}_0^{\bf a}(1 + \epsilon\delta_{k,k_p})$ with
large values of the perturbation wave-number $k_p$ to generate de-localised
small-scale perturbations in the system. From this, we define the decorrelator as $|\hat\Delta_k|^2 =  \langle|\hat{v}_k^{\bf a} - {\hat
v}^{\bf b}_k|^2\rangle$ or, in physical space, $\Delta = v^{\bf a} - v^{\bf b}$. 

Since both fields {\bf a} and {\bf b} satisfy the Galerkin-truncated Burgers equation, we can write down the evolution 
equation:
\begin{equation}
\frac{\partial \Delta}{\partial t} + \pkg\left [\frac{\partial \Delta v^{\rm th}}{\partial x} + \frac{1}{2}\frac{\partial \Delta^2}{\partial x}\right ] = 0;
\label{eps-equation}
\end{equation}
with initial conditions, most conveniently defined in Fourier space, as 
$\hat{\Delta}_k(t=0) = \epsilon_0{\hat v}_0^{\rm th}\delta_{k,k_p}$ and the projector $\pkg$ 
constraining the dynamics on a finite dimensional subspace with a maximum wavenumber $\kg$ and $N_G$ degrees 
of freedom. At
\textit{short} times, we linearise (for the same reasons as outlined for the 3D thermalised fluid) by 
dropping the quadratic non-linearity of $\Delta$ and obtain 
estimates, made precise in the next section, of an exponential, $k$-independent growth of 
$|\hat{\Delta}_k|^2$ consistent with our findings for the Euler equation. In the main paper, Fig. 3 (inset) 
shows plots of the decorrelator obtained from our linear theory; the agreement in the exponential phase 
with the decorrelator obtained from the full DNSs is remarkable. However, as with the 3D thermalised fluid, 
the approximations which go into the linear theory---dropping of the quadratic term---fails at later times. Hence, 
while the actual decorrelator measured from simulations of the full nonlinear system saturates, the one 
obtained from the linear theory continue to grow exponentially.

\section{Appendix D: Decorrelators: Bound on the Lyapunov Exponent}

The linear theory developed above for the 3D and 1D fluids are not just as useful to predict the nature of decorrelators at early times, 
but they are indispensable to estimate the Lyapunov exponents and their dependence on both temperature $T$ and degrees of freedom 
of the system $N_G$. 

For the 3D thermalised fluid, the linearised theory as summarised in
Eq.~\ref{strainCouplingP31} leads to the following bound on the growth of the
decorrelator $\dot{\Phi}(t) \le  -2 \overline{\gamma}_3 \Phi(t)$ and hence the
Lyapunov exponent $\lambda \le  -2\overline{\gamma}_3$.  As we show in
the main paper (Fig. 5, upper inset), results from our DNSs confirms this bound
as we find $\lambda \approx -0.62\overline{\gamma}_3$.

In order to uncover the dependence of $\lambda$ on $T$ and $N_G$, we exploit the fact that the 
linear theory helps us to associate the Lyapunov exponent with the eigenvalues of the strain tensor. 
Hence, the statistics of this tensor, which depends only on the properties of the velocity field of the
thermalised fluid determines the functional form of $\lambda$.

For notational simplicity, we denote $\hat{v}_i(\mathbf{k})$ as the Fourier components of the thermalised fluid 
and estimate: 
\begin{eqnarray} \left\langle \mathrm{Tr}\left[ \mathbf{S}^2 \right] \right\rangle
	= \left\langle S_{ij}S_{ij} \right\rangle &\approx& \displaystyle
	\sum_{\mathbf{k},\mathbf{k}^{\prime}}^{}\:-k_i k_j^{\prime}\left\langle
	\hat{v}_i(\mathbf{k})\hat{v}_j(\mathbf{k}^{\prime}) \right\rangle \nonumber \\
	&\approx& \dfrac{E}{N_{G}^{3}}\displaystyle \sum_{\mathbf{k}}^{}\:k^2 \approx E
N_{G}^{2} \label{eigenvalueDependence} \end{eqnarray}
leading to $\lambda \sim N_G \sqrt{T}$. 

For the 1D problem, a similar estimate is obtained by simpler manipulations of the linearised evolution equation for the decorrelator 
$|\hat{\Delta}_k|^2$: 

\begin{equation} 
\frac{\partial |\hat\Delta_k|^2}{\partial t} -
\frac{\sqrt{T}}{N_G}\sum_{q=1}^{N_G}q\left[\hat{\Delta}_q\hat{\Delta}_{-k}
+ \hat{\Delta}_{-k}\hat{\Delta}_{-k-q}\right] = 0.  
\label{simple-ek}
\end{equation} 
We have confirmed, numerically, that at short times
$\hat{\Delta}_{q}\hat{\Delta}_k$ remains spectrally flat, i.e.,
$\hat{\Delta}_{q}\hat{\Delta}_k \propto |\Delta_k|^2$, up to some undetermined
numerical constant.  Hence, and by using the identity $\sum_{q=1}^{N_G}q
\approx N_G^2$ (for large $N_G$), we obtain (where $C$ is a numerical constant) 
$|\Delta_k|^2 \propto e^{CN_G\sqrt{T}t}$ and thus, just like for the 3D thermalised fluid, $\lambda \sim N_G\sqrt{T}$. 

In the main paper, Fig. 4 has plots of the Lyapunov exponents from our DNSs for both 1D and 3D thermalised fluids which 
confirms the validity of our theoretical estimate.

\section{Appendix E: Decorrelators -- Saturation}

While we do understand why and at what time scales $t_{\mathrm{sat} }$ the decorrelators of thermalised fluids saturate (Fig. 3), 
it still remains to be understood how they approach the saturated value. To understand this for the 3D thermalised fluid, for simplicity, 
we define a normalised decorelator $\overline{\Phi}(t)=\frac{\Phi(t)}{2E}$. Given that the only time scale in the problem is the inverse 
of the Lyapunov exponent, we construct the following empirical form :
\begin{align}
  \overline{\Phi}^{\mathrm{emp}}(t) &= \left( 1+\mathrm{exp}\left[-\lambda (t-t_{\mathrm{sat}})\right] \right)^{-1}
	\label{saturationEmpirical}
\end{align}
with a saturation time-scale $t_{\mathrm{sat}} \sim 1/\lambda$ but found more precisely by fitting the data from our simulations. 
In Fig.~\ref{fig:SI}(c) we show a representative plot illustrating how the empirical form approximately fits the data. 

While the functional form of the decorrelator $\overline{\Phi}(t)$ defined above is purely heuristic it does serve 
to underline the fact that the nature of many-body chaos is determined solely by the Lyapunov exponent.\\ 

\section{Appendix F: Decorrelators -- Spatial Spread}

\textit{Non-locality} is inherent in 3D thermalised flows due to the pressure term as well as Galerkin truncation.
Hence it allows the perturbation seeded locally at $t=0$ to affect the evolution of thermalised velocity everywhere. 
This is already seen in Eq.~\ref{evolutionPerturbation} which shows that at $t=0^+$, at spatial points far from the center of perturbation, 
the growth of $\bm{\delta }\mathbf{v}(\mathbf{x})$ is essentially triggered by the \textit{non-local} integral  term.
The subsequent growth of the difference field is then through its coupling 
with the rate-of-strain tensor.

All of this suggest that the spatially resolved decorrelator $\phi(r,t)$ will not have a \textit{wavefront} which propagates (radially) 
with a finite \textit{butterfly} speed. On the contrary, as was also suggested in the inset of Fig. 2(a) in the main paper, 
one should expect a self-similar spatial profile for decorrelator, i.e.,  $\phi(r,t) \sim r^{-\alpha}$.

In Fig.~\ref{fig:SI}(d), we see clear evidence of $\phi(r,t) \sim r^{-\alpha}$, with $\alpha \approxeq 4$, for in the range 
$0 \ll r \ll \pi$ (where $\pi$ is half the system size since the perturbation is seeded in the middle of a 2$\pi^3$ cubic box). 
A further consequence of this (Fig.~\ref{fig:SI}(d), inset) is that the isocontours of the decorrelator (measured through a suitable 
threshold value  $\phi_0$) are spread in space-time as $t\sim \ln r$. 

While we do not have a way of obtaining the exponent $\alpha \approxeq 4$ analytically, the constraint that 
$\Phi(t) = \displaystyle \int_{0}^{L}dr\: r^2\phi(r,t)$ must be bounded (from above) suggests that $\alpha > 3$ which 
is consistent with what we measure in our data.

Given that for the 1D Burgers problem, we carry out the analysis entirely in Fourier space, the seed perturbation 
is also introduced in Fourier space and hence not localised in physical space. Therefore the question of the spatial spread of 
decorrelators remains unanswered for 1D thermalised fluids in this study.

 \bibliography{referencesMain}

\end{document}